\documentclass[a4paper,twocolumn]{article}
\usepackage{graphicx}
\usepackage{amsmath}
\usepackage{txfonts}
\usepackage[pdftex,bookmarksnumbered=true,breaklinks=true]{hyperref}
\usepackage{bbold}
\usepackage{color}
\usepackage{cite}
\usepackage{authblk}
\usepackage{abstract}

\begin{document}

\title{In-Situ Dual-Port Polarization Contrast Imaging of Faraday Rotation in a High Optical Depth Ultracold $\mathrm{^{87}Rb}$ Atomic Ensemble}

\author[1]{Franziska Kaminski \thanks{kaminski@nbi.dk}}
\author[1]{Nir S. Kampel}
\author[1]{Mads P. H. Steenstrup}
\author[1,2]{Axel Griesmaier}
\author[1]{Eugene S. Polzik} 
\author[1]{J\"org H. M\"uller \thanks{muller@nbi.dk}}
                     
\affil[1]{
Niels-Bohr-Institute, Danish Quantum Optics Center QUANTOP, Copenhagen University, Blegdamsvej 17, 2100 Copenhagen, Denmark}
\affil[2]{
Physikalisches Institut, Universit\"at Stuttgart, Pfaffenwaldring 57, 70569 Stuttgart, Germany}

\renewcommand\Authands{ and }

\date{\today}

\twocolumn[
\maketitle
\begin{onecolabstract}
We study the effects of high optical depth and density on the performance of a light-atom quantum interface. 
An in-situ imaging method, a dual-port polarization contrast technique, is presented. This technique is able to compensate for image distortions due to refraction. We propose our imaging method as a tool to characterize atomic ensembles for high capacity spatial multimode quantum memories.
Ultracold dense inhomogeneous Rubidium samples are imaged and we find a resonant optical depth as high as 680 on the D1 line.
The measurements are compared with light-atom interaction models based on Maxwell-Bloch equations. We find that an independent atom assumption is insufficient to explain our data and present corrections due to resonant dipole-dipole interactions.
\vspace{0.03\textheight}
\end{onecolabstract}
]
\saythanks

\section{Introduction}

The storage and retrieval of single photons \cite{Duan:2001tt} and continuous variable quantum states \cite{Julsgaard:2004gp} in quantum memories \cite{Zhang:2009fi,Simon:2010kl} has become a major endeavor for the realization of quantum networks \cite{Kimble:2008if}. While single-qubit memories are sufficient to establish a secure communication channel, being able to store more qubits increases the capacity of the channel. It has been shown that not only the fidelity of storage and read-out but also the multimode capacity of an ensemble scales favorably with the on-resonant optical depth (OD) \cite{Zeuthen:2011iv}.

Spatially resolved detection is a requirement for multimode memories \cite{Vasilyev:2010gx} that store each polarization qubit in an independent spatial light mode.

Polarization rotation, also called Faraday rotation, is a well known effect \cite{Budker:2002ii}. It has been a means to generate an entangled state of two atomic ensembles \cite{Julsgaard:2001tu}, to measure the OD of atomic ensembles \cite{Kubasik:2009be,Takahashi:1999ib}, spin dynamics \cite{Smith:2003ue,Liu:2009ff} and magnetic fields \cite{Terraciano:2008cm} and was used as a gigahertz-bandwidth probe \cite{Siddons:2009kw}.

When applying the Faraday technique to dense, inhomogeneous, high OD samples new challenges arise.
A large OD, together with the sample inhomogeneity, leads to stronger refraction or lensing, distorting images. A small transverse size leads to diffraction, posing stringent constraints on the properties of the imaging system. A large density leads to effects beyond the independent atom hypothesis commonly applied in quantum optics \cite{Rath:2010cz,Weller:2011fo}.

We present an imaging method that reduces distortions due to refraction and we introduce a model that treats the influence of light assisted cold collisions on dispersive interactions.

\begin{figure}
\includegraphics[width=0.5\textwidth]
{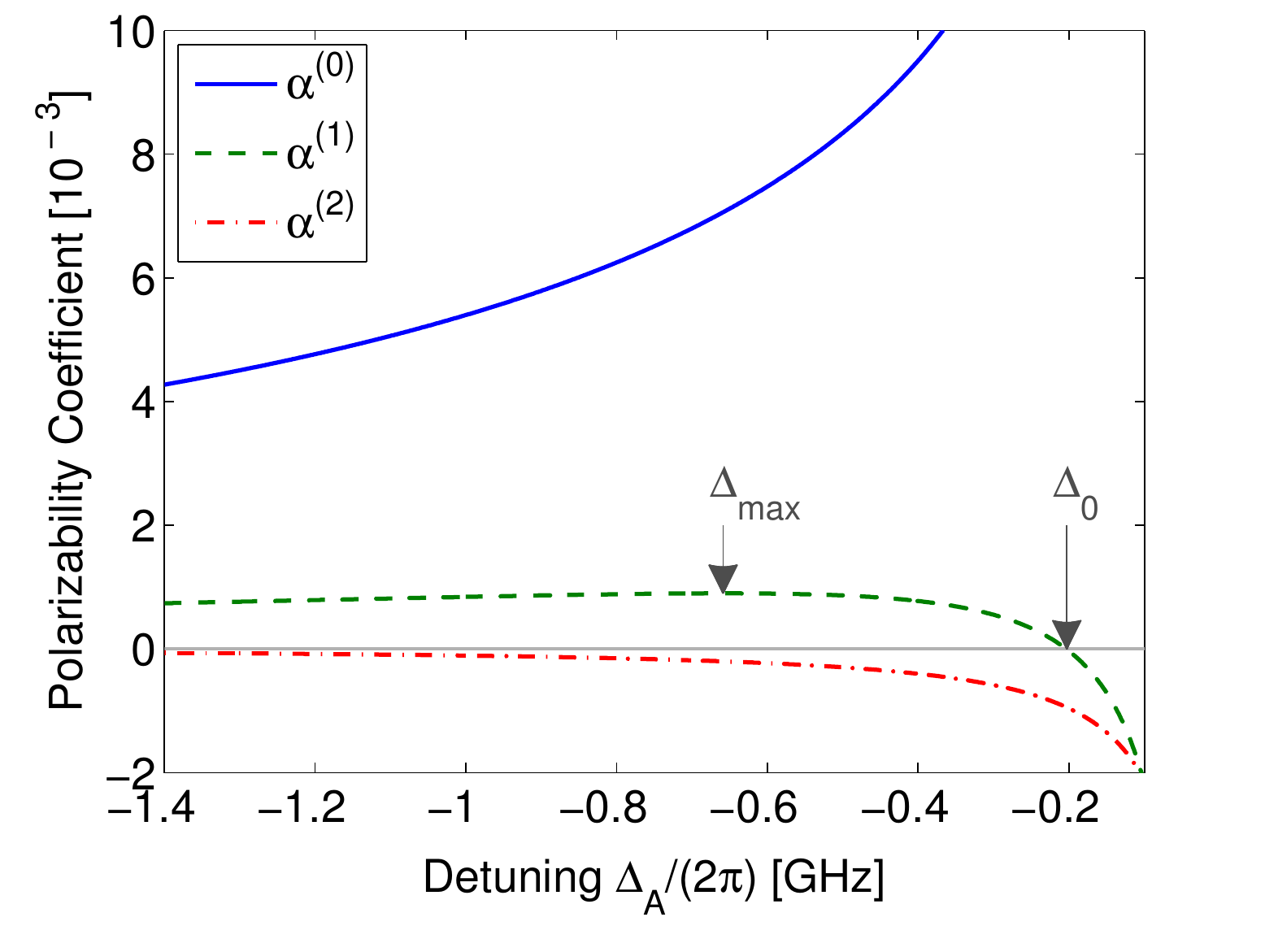}%
\caption{
Polarizability coefficients normalized to be unit free by $2 |\langle J||d||J'\rangle |^2/\hbar\Gamma_A$. $\alpha^{(0)}$ describes the scalar refractive index, $\alpha^{(1)}$ is responsible for Faraday rotation and $\alpha^{(2)}$ is responsible for ellipticity changes and the corresponding Raman population transfers. The zero crossing of the $\alpha^{(1)}$ coefficient is indicated as $\Delta_0$ and its maximum as $\Delta_{max}$.
}
\label{fig:polarizability}
\end{figure}

There are already several dispersive imaging techniques available.  Phase-contrast imaging \cite{Meppelink:2010hl, Higbie:2005bi} uses a spatially selective phase plate and is hardly polarization sensitive. Therefore it mainly measures the scalar part of the polarizability, $\alpha^{(0)}$, or scalar refractive index (Fig.~\ref{fig:polarizability}).
Single-port polarization-contrast imaging \cite{Bradley:1997ib} uses absorptive polarizers and detects the vector part, $\alpha^{(1)}$, or refractive index differences but does not cancel diffraction. 
Dark-ground imaging \cite{Andrews:1996ch} uses a spatially selective block and is sensitive to both $\alpha^{(0)}$ and $\alpha^{(1)}$. 
Another technique records directly the diffracted wave of an object and the image is reconstructed numerically \cite{Turner:2005et}.

We employ a polarization-contrast imaging method which uses a polarizing beam splitter (PBS) instead of an absorptive polarizer, enabling us to record the full light intensity (dual-port). This gives access to the scalar and vector components of the polarizability simultaneously and distinguishably. It enables us to cancel diffraction in the same way common-mode noise is canceled in a differential photodiode and leaves us sensitive to the spatial profile of the Faraday rotation signal. All the described methods are weakly sensitive to the tensor part of the polarizability, $\alpha^{(2)}$, leading to changes in the coherences and populations of the atoms during the interaction. This manifests in an excess ellipticity of the light. High sensitivity to the tensor components can be obtained with circular polarizers.

A Raman/beam splitter-type memory \cite{Hammerer:2010wf} is based on the $\alpha^{(2)}$ part of the polarizability making use of Raman population transfers and coherences. In our specific case of $^{87}$Rb it can be realized using the $m_F = \pm 1$ states of the  F=1 manifold (Fig.~\ref{fig:setup}). 
To avoid unwanted differential phase imprints it is advantageous to minimize the $\alpha^{(1)}$ part of the polarizability. On the D1 line $\alpha^{(1)}$ is expected to vanish at a detuning $\Delta_0/(2\pi)=-204 \mathrm{MHz}$ (Fig.~\ref{fig:polarizability}). A local maximum of $\alpha^{(1)}$ is expected for red detunings at $\Delta_{max}/(2\pi) = -660 \mathrm{MHz}$.

Scalar diffraction is due to the $\alpha^{(0)}$ part of the polarizability that is dominant over the whole range of explored detunings. An imaging method that can address these distortions is therefore desirable.

Dual-port polarization contrast imaging, also referred to as Faraday imaging in this paper, can be more generally employed for spatially resolved detection of the atomic magnetization and the study of quantum coherence in degenerate gases and solid state systems \cite{Higbie:2005bi,Vengalattore:2010hz}.

\section{Experimental Setup}
\label{sec:setup}

\begin{figure}
\includegraphics[width=0.5\textwidth]
{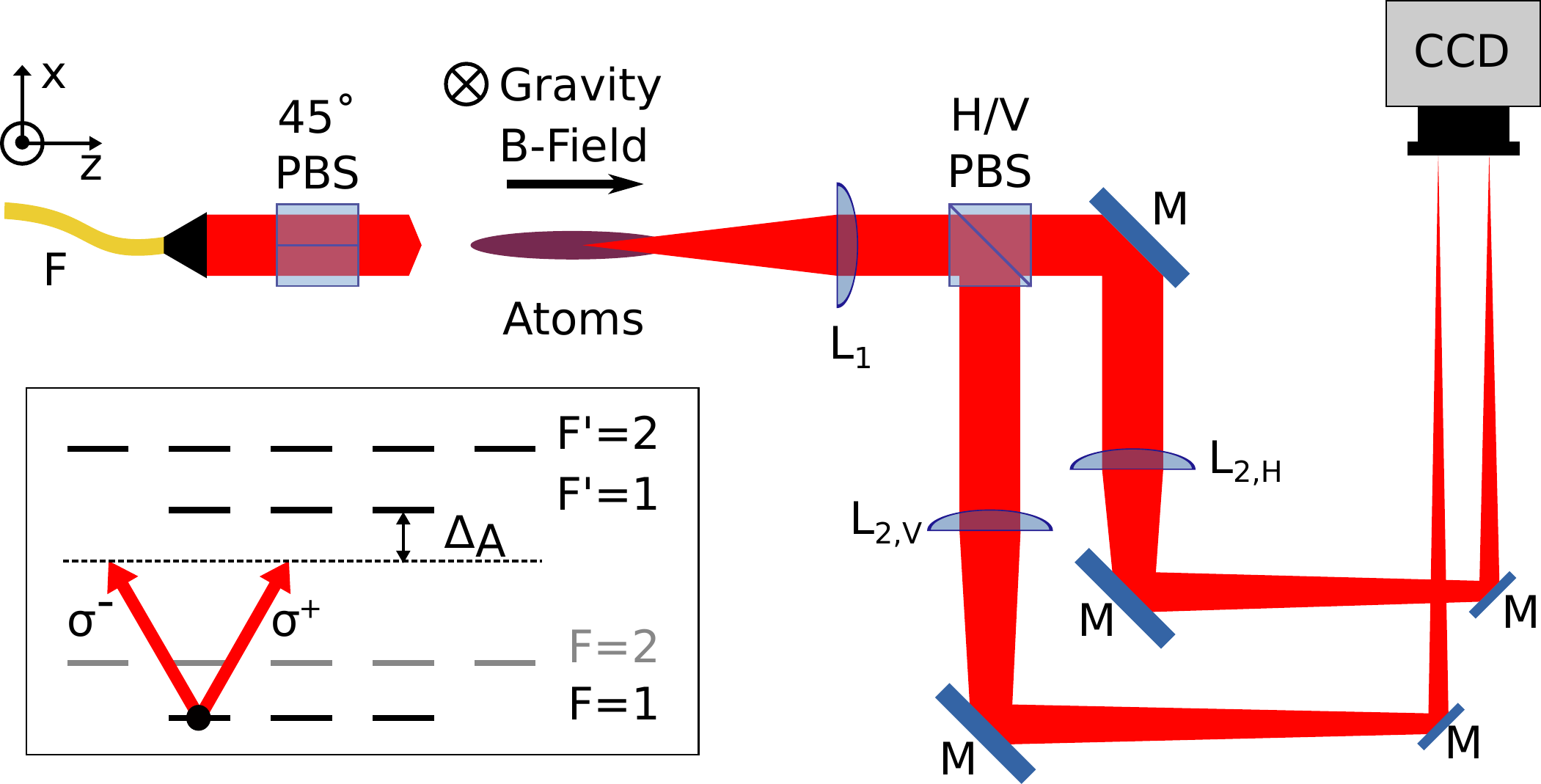}%
\caption{Optical setup showing the $45^{\circ}$ PBS preparing the probe light after the Fiber (F), the placement of lenses $L_1$, $L_{2,H}$ and $L_{2,V}$, mirrors (M) and analyzing cube H/V PBS as well as the CCD camera. The B-field direction is mainly along z. The inset shows the $^{87}$Rb D1 line level scheme and the linear probe light in the z quantization axis.}
\label{fig:setup}
\end{figure}

The ultracold thermal atomic ensemble is prepared in a Ioffe-Pritchard type magnetic trap 
with frequencies $\omega_r=2\pi \cdot (115.4 \pm 0.5) \,\mathrm{s}^{-1}$ and $\omega_z=2\pi \cdot (11.75 \pm 0.25) \,\mathrm{s}^{-1}$. The atoms are spin polarized in the $|F=1,m_F=-1 \rangle$ state in the quantization axis defined by the local magnetic field. The main B-field component is oriented along the propagation direction of light (z). The gravitational sag of $18 \mathrm{\muup m}$ leaves the B-field at an angle between the z and gravity axis (-y) of about $15^{\circ}$ in the center of the trap. There is a $5^{\circ}$ variation within $w_r$ and a $0.25^{\circ}$ variation within $w_z$.

The first element in the optical assembly, Fig.~\ref{fig:setup}, is a polarizing beam splitter oriented at $45^{\circ}$ to prepare a clean linear polarization. The probe beam enters the ensemble with an $\exp(-2)$ beam radius of $140\mathrm{\muup m}$ with a flux of \linebreak $1100~\mathrm{photons/\muup s/\muup m}^2$ which corresponds to a spontaneous emission probability of 0.03 in the pulse duration at a detuning $\Delta_A/(2\pi)=-200\mathrm{MHz}$. The pulse duration is $10\muup$s. To average over shot-to-shot atom number fluctuations we identically prepare each experimental run five times.

After the atomic ensemble an achromatic lens doublet ($L_1$) collimates the scattered light, which is then split into horizontal (H) and vertical (V) polarization components by another PBS and imaged with identical lenses $L_{2,V/H}$ onto separate areas of a CCD camera. We take images $I_V$ and $I_H$ with atoms present, images $I_V^{\mathrm{ref}}$ and $I_H^{\mathrm{ref}}$ without atoms present and bias images to correct for any stray light and electronic offset. We balance the detection to split unscattered light to equal amounts into both output arms, $I_V^{\mathrm{ref}}=I_H^{\mathrm{ref}}$. Any small remaining imbalance is corrected for during post-processing.
This allows us to detect the difference $I_H-I_V$ and the sum $I_H+I_V$ simultaneously with optimal sensitivity. Centering the images $I_{H/V}$ on top of each other is critical for the accurate determination of Faraday angles. We therefore apply a fitting algorithm on images at large detuning and the determined position of the images is then used for the whole data set. We achieve sub-pixel resolution for the centering.

The Faraday angle $\theta_F$ is the rotation angle of linearly polarized light. We deduce the angle on each pixel by:

\begin{equation}
\theta_F=\frac{1}{2} \arcsin \left( \frac{I_H-I_V}{I_H+I_V} \right).
\end{equation}

This will give an accurate polarization rotation angle, if we can neglect the presence of any circular polarizations after the interaction. The sensitivity of our method is illustrated by our ability to detect a small rotation of $0.038^{\circ}$, produced by the cell windows that are subject to large magnetic fields due to the magnetic trap, for which we correct the data.

We adjust $L_1$ to image the end plane of the ensemble. The diffraction limited imaging resolution on the object plane is $3.6\mathrm{\muup m}$. The magnifications with 95\% confidence bound in both arms of our imaging system are $12.83 \pm 0.08$ and $12.87 \pm 0.08$, i.e. identical within measurement accuracy. Images of clouds falling freely under the influence of gravity are used to determine the magnification. The acceptance full opening angle of the system is $15.2^{\circ}$.

For quantitative imaging the opening angle of the imaging system needs to be sufficiently large compared to diffraction angles of the ensemble. 
We estimate the full geometric diffraction angle of the ensemble is $\alpha_G \approx \lambda/4d < 1^{\circ}$, where $\lambda$ is the wavelength and d the radial extent of the ensemble. The angle due to refraction or lensing can be approximated by comparing the phase shift $\phi$ in the center to the one at the edge of the sample $\alpha_L=2(\phi(r=0)-\phi(r=d)) \lambda/(\pi d)$ \cite{Andrews:1996ch} and reaches at $\Delta_A/(2\pi)=-200\mathrm{MHz}$ $\alpha_L(\Delta_0) \approx 0.3^{\circ}$. 
Both are sufficiently small to neglect light loss at the aperture of the imaging system. 
We compared these estimates to more elaborate diffraction simulations \cite{Muller:2005jr,Zeuthen:2011iv} including a model of the full imaging system. These confirmed the conclusions of the simple diffraction estimates.

As an independent sample characterization we perform standard absorption imaging on the D2 line
after a time of flight (TOF) of 45 ms. Using the magnification $1.577 \pm 0.002$ and independently measured trap frequencies we can specify the waists of the in-trap density distribution \linebreak $\rho(r,z) = \rho_0 \exp( -r^2 / (2 w_r^2) - z^2 / (2 w_z^2))$ as $w_r=7.7 \mathrm{\muup m}$ and $w_z=70.5 \mathrm{\muup m}$, leading to a Fresnel number of $F = w_r^2 / (2 w_z \lambda ) = 0.5$. From these parameters we determine the temperature $k_B T_i = M w_i^2 \omega_i^2$ and $T = (T_r^2T_z)^{1/3}=300$nK, where M is the atomic mass. 

By using the D2 line scattering cross section for the cycling transition $\sigma_{D2} = 3\lambda_{D2}^2/2\pi$ we determine a peak density of $\rho_0^{\mathrm{abs}} = 1.2 \cdot 10^{19} \mathrm{m}^{-3}$ and an atom number of  $N_{\mathrm{at}}^{\mathrm{abs}} = 8.1 \cdot 10^5$ as an average over all data points.

Both absorption imaging and Faraday measurements allow to deduce the number of atoms using models for the optical cross sections. Comparing the deduced numbers and taking into account the thermodynamic properties of ultracold gases allows to identify systematic errors in either method. While absorption imaging is a standard method, it is well known that it is difficult to estimate the precise effective scattering cross section due to uncertainties in magnetic field alignment, light polarization quality and repump efficiency\footnote{In later control experiments we found a systematic undercount of atoms by a factor of two in absorption imaging due to insufficient repumping.}
\cite{Gerbier:2004go}. 
Since the cycling transition allows for the maximal cross section $\sigma_{D2}$ our measured $N_{\mathrm{at}}^{\mathrm{abs}}$ and $\rho_0^{\mathrm{abs}}$ are hard lower bounds.
We estimate a hard upper bound by noting that we do not observe condensed atoms on the absorption images and hence $T/T_c > 1$. The condensation temperature of an ideal gas is $k_B T_c = \hbar \bar{\omega} (N_{at}/\zeta(3))^{1/3}$, with $\bar{\omega}=(\omega_r^2\omega_z)^{1/3}$ and $\zeta(3)$ is the Riemann zeta-function. We correct $T_c$ for the effects of finite size and mean field interactions. Both effects reduce $T_c$ for our parameters by maximal 1.2\% and 5.5\% respectively  \cite{Gerbier:2004go,Dalfovo:1999vn}. Corrections to the determined ensemble temperature T arise from the use of a Gaussian instead of a Bose distribution in the fit model. We estimate this systematic correction by fitting Gaussian profiles to analytical Bose enhanced densities and find a systematic underestimation of the temperature by about 10\%. 

Taking all these corrections into account we reach $T/T_c=1$ for several single-run data points at an atom number scaling factor $f=N_{at}/N_{at}^{\mathrm{abs}}$ of maximal $f_{max}=2.7$, defining a hard upper bound for the real atom number $N_{at}$. We will show below that this upper bound is still too low to allow the Faraday rotation data to be fitted with an independent atom model and we conclude that lineshape corrections due to light assisted collision become relevant.

\section{Experimental Results}

\begin{figure}
\includegraphics[width=0.5\textwidth]{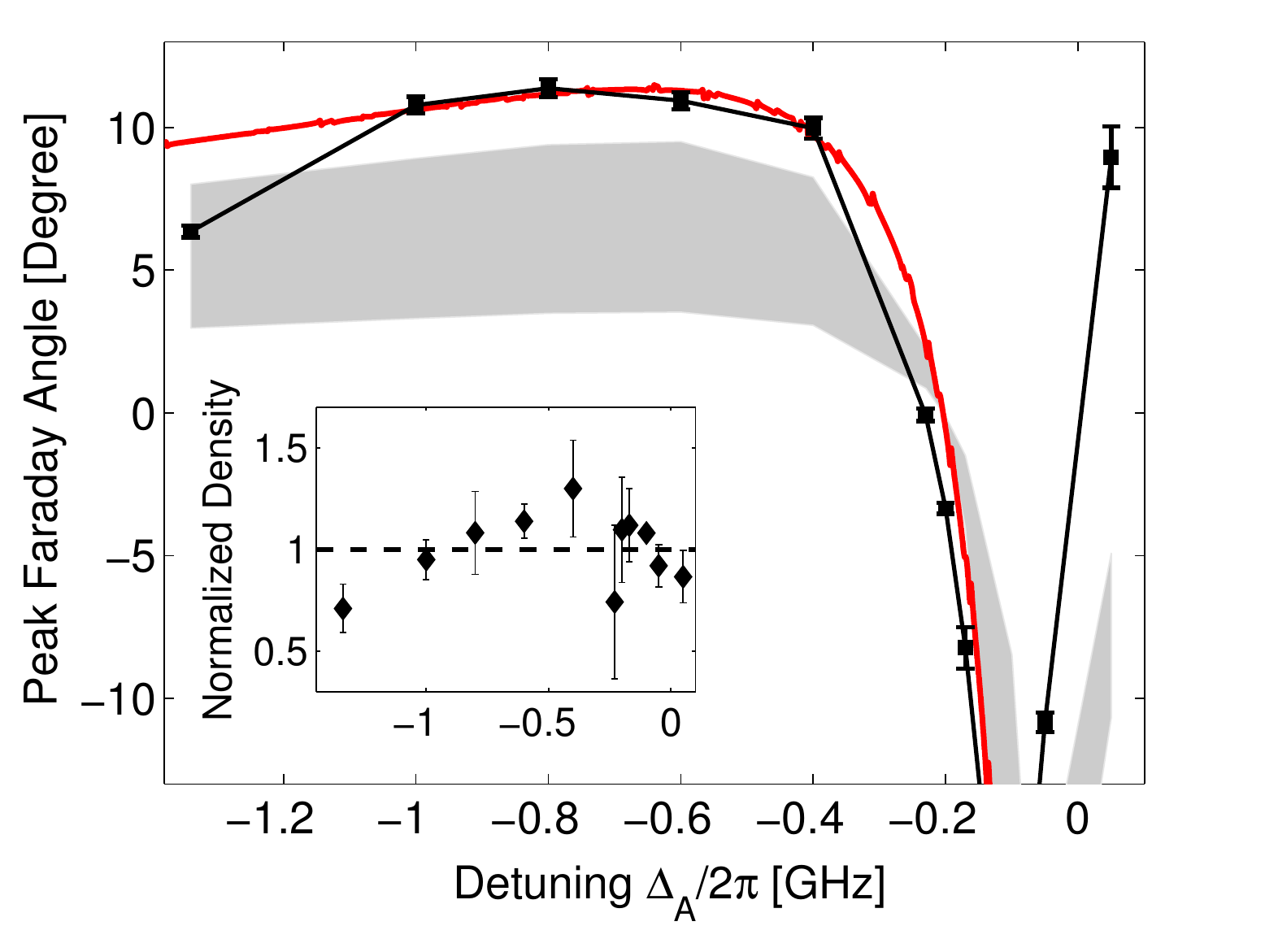}%
\caption{Detuning dependence of peak Faraday angle. Experimental data (black squares) is best reproduced by a model including light assisted cold collisions (red solid line). The grey area shows the prediction of a coupled Maxwell-Bloch model assuming independent atoms for the permissible range of the atom number scaling factor $f$ (see text). The inset shows the relative atomic density variation with detuning.}
\label{fig:fdetuning}
\end{figure}

Figure~\ref{fig:fdetuning} shows the detuning dependence of the observed peak Faraday angles (black squares) together with a model including light assisted cold collisions (red line) and a coupled Maxwell-Bloch model that assumes independent atoms (grey area). Both models are discussed in Sec. \ref{sec:models}. 
Both use input parameters deduced from absorption imaging, i.e. the sample radii $w_r$ and $w_z$ and the atom number $f N_{at}^{abs}$, averaged over all detunings. We infer an optimal atom number scaling $f$, by matching the light assisted cold collision model to the experimental data and obtain $f_{opt}=2.13$ (\footnote{This corresponds to a temperature ratio of $T/T_c=1.27$.}). The grey shaded area indicates the atom number scaling range $1<f<2.7$, defined in Sec. \ref{sec:setup}.
Experimental peak angles are determined by averaging over 3x3 pixels around the determined center positions of the density distribution and the error bars are the standard deviation of 5 experimental runs.
The small structure on the red line is due to vibrational molecular resonances, discussed further below. 
The figure inset shows the variation of densities with detuning normalized to the averaged density entering the models.
The discrepancy between the data point at $\Delta_A/(2\pi)=-1340\mathrm{MHz}$ and the models might be explained by the low density.

\begin{figure}
\includegraphics[width=0.5\textwidth]{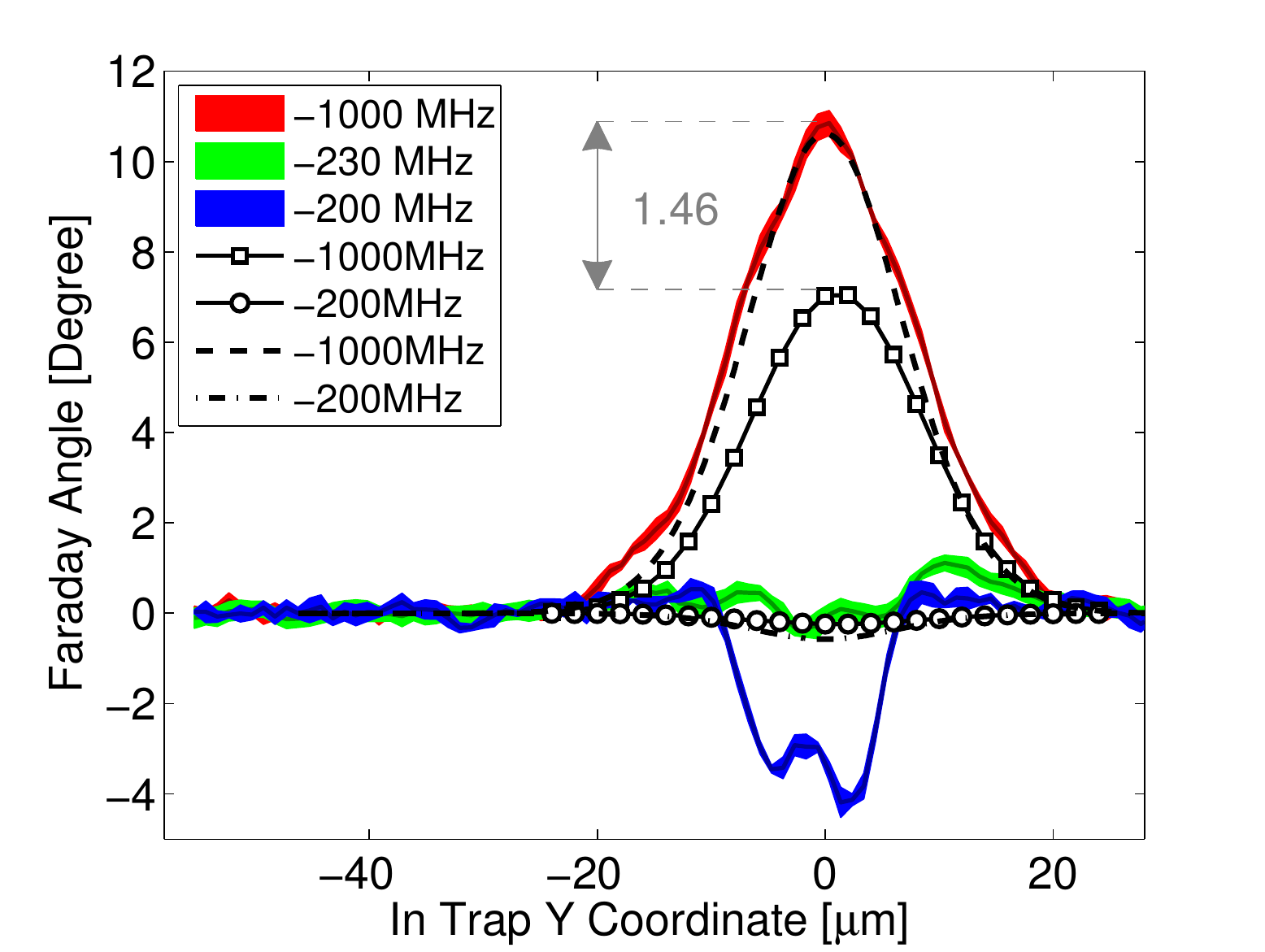}%
\caption{Cut through Faraday angle images for various detunings $\Delta_A$. Colored areas represent the standard deviation of 5 experimental realizations. Coupled Maxwell-Bloch simulations assuming independent atoms (squares, circles) underestimate the angle by a factor 1.46, compared to a model including light assisted cold collisions (dashed, dot-dashed).
}
\label{fig:ftrace}
\end{figure}

Figure~\ref{fig:ftrace} shows the spatially resolved Faraday angle as deduced from the camera images for the detunings $\Delta_A/(2\pi)=-\{1000,230,200\}\,\mathrm{MHz}$ averaged over five realizations together with the two model predictions for input atom number $f_{opt}N_{\mathrm{at}}^{\mathrm{abs}}=1.73 \cdot 10^6$.
The good reproducibility of sample preparation is evidenced by the small standard deviation encoded in the colored areas around the averaged profiles.
The spatial shape of the Faraday angle profile at $\Delta_A/(2\pi)=-1000\mathrm{MHz}$ (red) fits the expected shape from the light assisted cold collision model when the finite imaging resolution is taken into account.
The experimental profile at $\Delta_A/(2\pi)=-200\mathrm{MHz}$ deviates significantly from the shape of the density distribution. 
We observe minimal Faraday rotation at $\Delta_A/(2\pi)=-230\mathrm{MHz}$, shifted by about $30\mathrm{MHz}$ from the expected detuning $\Delta_0$. This shift is outside possible systematic errors in the frequency scale.

\begin{figure}
\includegraphics[width=0.5\textwidth]{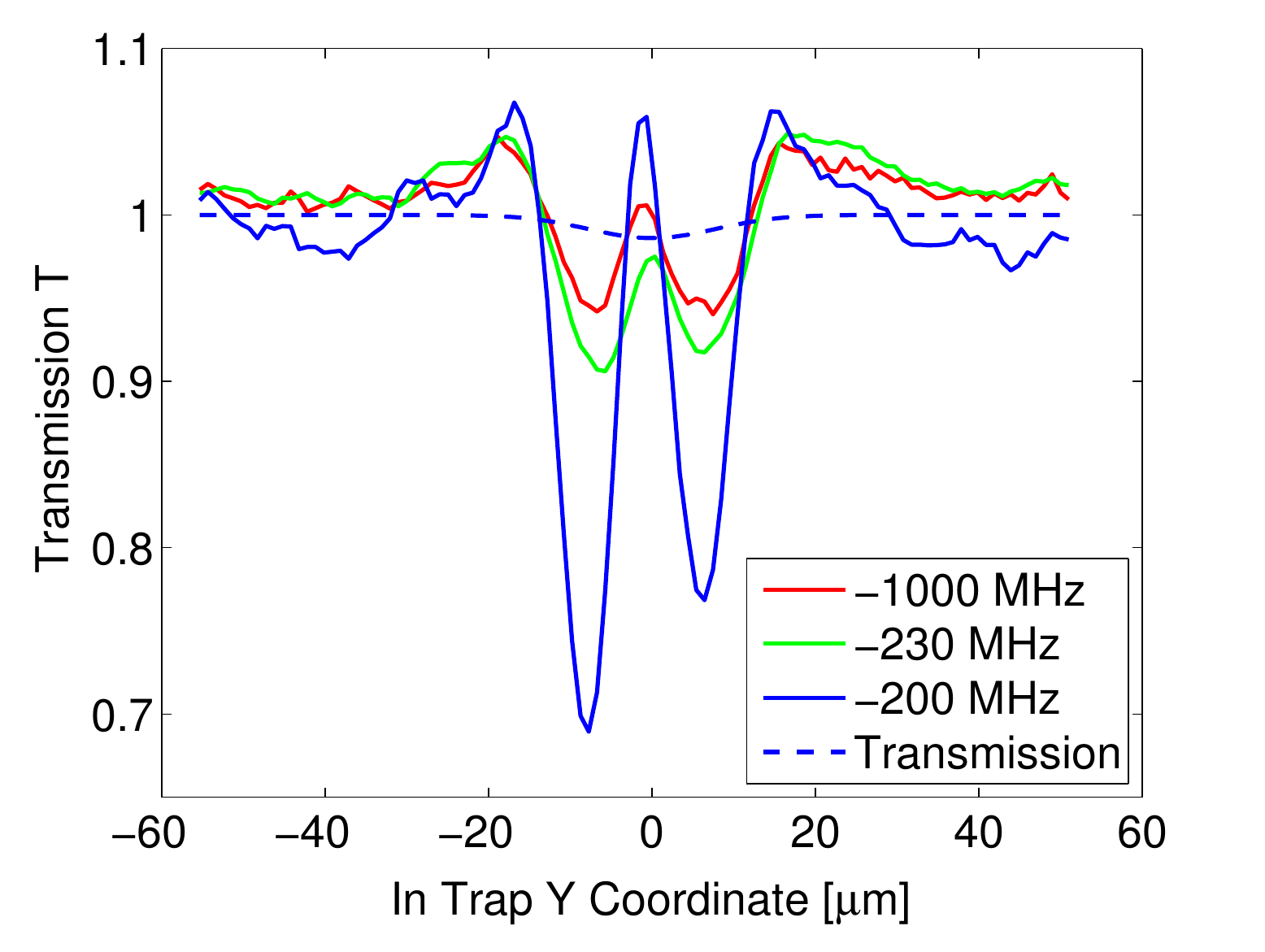}%
\caption{
Spatially resolved transmission: cut through $T=(I_H+I_V)/(I_H^{\mathrm{ref}}+I_V^{\mathrm{ref}})$ for various detunings compared to the column density approximated transmission profile of the ensemble at $\Delta_A/(2\pi)=-200\mathrm{MHz}$ (dashed). Diffraction dominates over absorption for all detunings.
}
\label{fig:transmission}
\end{figure}

Figure~\ref{fig:transmission} shows a three pixel averaged cut through the transmission $T=(I_H+I_V)/(I_H^{\mathrm{ref}}+I_V^{\mathrm{ref}})$, where we normalize with the reference images. This allows us to visualize the effect of intensity redistribution across the image due to refraction and diffraction.
To indicate the expected photon loss we plot an estimated transmission profile for $\Delta_A/(2\pi)=-200\mathrm{MHz}$ using a naive column density model neglecting diffraction effects.
The expected photon loss is hardly distinguishable from the detection noise.
From the spatial transmission curves it is apparent that data is dominated by refraction rather than absorption for all detuning values shown.
We emphasize that due to the dual-port detection the distortion effects of diffraction are largely canceled in Faraday angle profiles.

\begin{figure}
\includegraphics[width=0.48\textwidth]{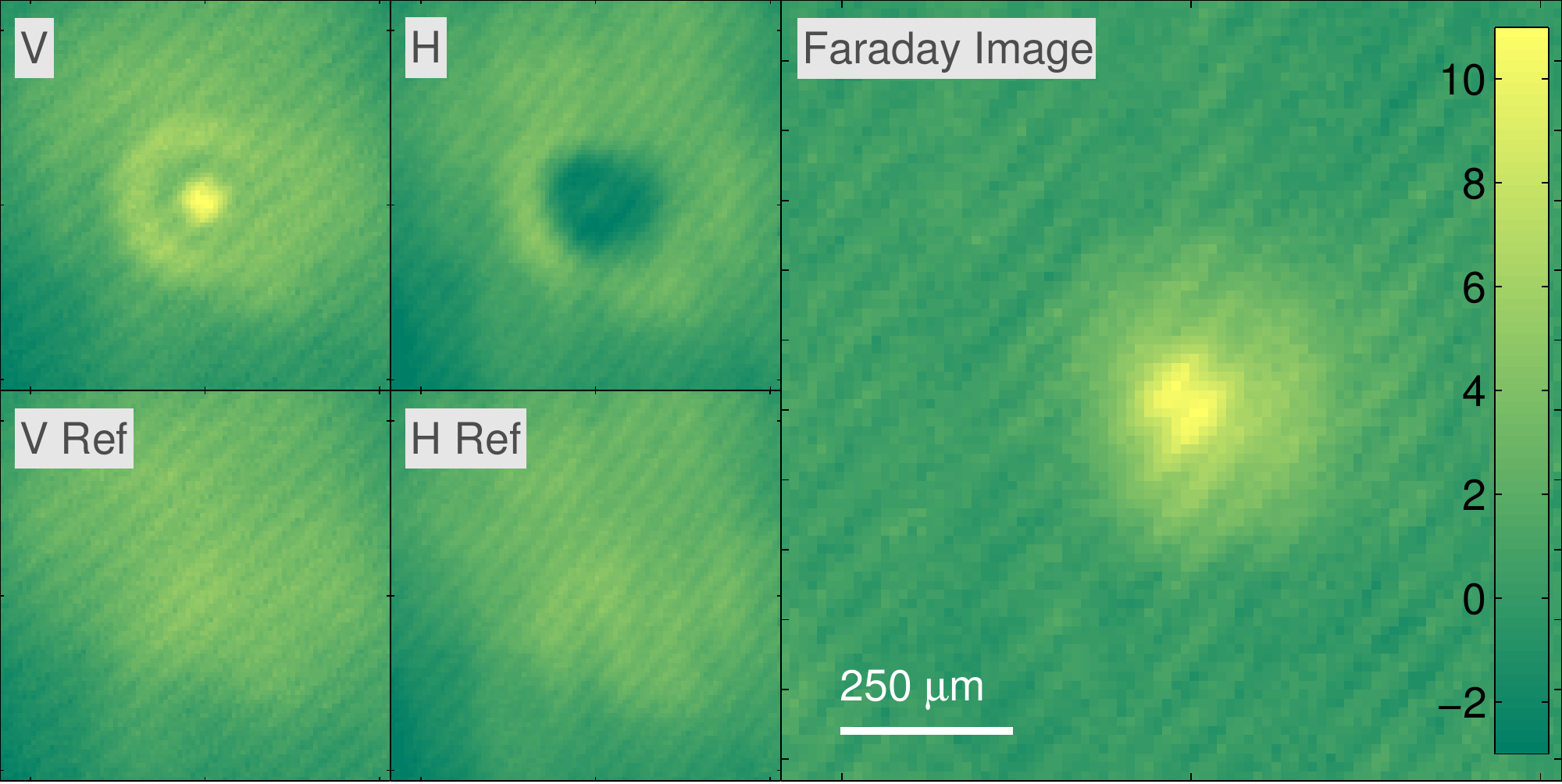}%
\caption{Raw images (left panel) and deduced Faraday image (right panel) for a detuning of $\Delta_A/(2\pi)=-400\mathrm{MHz}$. V/H refers to images with atoms present, while V/H Ref to images without atoms. The color scale indicates the Faraday angle in degree. The scale bar indicates dimensions on the camera. Diffraction rings on raw images V/H disappear on the Faraday image.}
\label{fig:images}
\end{figure}

This compensation of refraction effects is illustrated in Fig.~\ref{fig:images}, where raw images $I_H$, $I_V$, $I_H^{\mathrm{ref}}$ and $I_V^{\mathrm{ref}}$ are shown together with the 2-D reconstruction of Faraday rotation angles measured at a detuning of $\Delta_A/(2\pi)=-400\mathrm{MHz}$.

%
\section{Models}
\label{sec:models}

Our first model accounts for the collective response of all atoms, while treating each atom as an independent scatterer. The model is based on coupled Maxwell-Bloch (MB) equations  \cite{Kupriyanov:2005hf,Geremia:2006dv,Hammerer:2010wf} with excited states eliminated adiabatically, using continuous variables and including absorption. The probe light is propagated spatially through the ensemble while simultaneously evolving the ground state F=1 manifold populations and coherences in time. 
To incorporate the spatial inhomogeneity of sample density, initial atomic state and magnetic field, we extend the 1+1 dimensional geometry to 3+1 dimensions, following \cite{Mitchell:2009vd}. In the model light is assumed to propagate along straight lines and atomic motion is neglected.

The model Hamiltonian $\mathcal{H} = \mathcal{H}_{int}^{(0)} + \mathcal{H}_{int}^{(1)} + \mathcal{H}_{int}^{(2)} + \mathcal{H}_B$ contains the atom-light interaction decomposed into its irreducible tensor components $\mathcal{H}_{int}^{(j)}$ (App. \ref{sec:lightmatter}) and the effect of the external magnetic field  $\mathcal{H}_B = \vec{\Omega}(r) \cdot \vec{F}$. Here $\vec{\Omega}(r)$ is the the vector of Larmor frequencies and $\vec{F}$ is the total atomic angular momentum vector.
To compare the simulation results to our image data we plot the Faraday angle, time averaged over the probe pulse duration, at the output end of the atomic sample (Fig.~\ref{fig:fdetuning} and \ref{fig:ftrace}).

We can use the full MB model to quantify the combined effects of tensor polarizability and B-field inhomogeneity by comparison to a much simpler, idealized Faraday model. 

The idealized Faraday model starts from an effective two-level system description extended to our multilevel situation and neglects atomic evolution due to rank 2 tensor components and the magnetic field. 
The Faraday angle in this model is half of the accumulated phase difference between the left and right circularly polarized light.
The phase shift $\phi$ for a single polarization component is given by $\phi = 2\pi \lambda^{-1} \int n \mathrm{d}z$, where n is the refractive index and $\lambda$ is the wavelength.
Summing over all excited states ($i$) we obtain a Faraday angle

\begin{equation}
\theta_F^{A}(\Delta_A) = \frac{1}{2} \sum_i p^{(i)} \frac{\Delta_A^{(i)}}{\Gamma_{A}} \sigma^{(i)}_A(\Delta_A^{(i)}) \int \rho(\vec{r}) \mathrm{d}z,
\label{eq:FarAt}
\end{equation}

where $p^{(i)}$ is $+1$ for right-handed and $-1$ for left-handed circular polarization, $ \rho(\vec{r})$ is the atomic density, $\Gamma_{A}$ is the full atomic line width and $\Delta_A^{(i)}$ is the detuning to the excited state $i$.
The scattering cross section is $\sigma^{(i)}_{A}(\Delta_A^{(i)}) = \sigma_0^{(i)} / (1+4(\Delta_A^{(i)}/\Gamma_A)^2)$ with $\sigma_0^{(i)} = \xi_i^2 \cdot 3\lambda^2/(2\pi) \cdot  (2J'+1)/(2J+1) $ the on-resonant scattering cross section.
Here $J$ and $J'$ denote the total electronic angular momentum of the ground and excited state respectively. The parameter $\xi_i$ is defined via the transition dipole moment $d_i =\xi_i \langle J||d||J'\rangle$, such that in a two level system $\xi=1$. 
Comparing this simpler model to the full Hamiltonian dynamics we find that the effects of inhomogeneous magnetic field and tensor dynamics lead to a reduction of the Faraday angle by a constant factor $\beta_B=0.86$ in the range of detunings $\Delta_A/(2\pi)=-1340\mathrm{MHz}$ to $-400\mathrm{MHz}$.

Our second model addresses the effect of light assisted cold collisions. At high atomic densities atoms can no longer be treated as independent scatterers. Electronic energy levels for close pairs of atoms split and shift. The light scattering properties of a pair are modified compared to isolated atoms. This effect of the dipole-dipole interaction can be described by established methods from molecular physics \cite{Movre:1980tf,Movre:1999jn}.
We consider repulsive and attractive molecular potentials for ground-excited state Rb$_2^*$ atom pairs, neglecting hyperfine recoupling \cite{Kemmann:2004gz}, and calculate allowed energy levels. For the attractive molecular potentials the position of photoassociation resonances are calculated using the LeRoy-Bernstein formula \cite{LeRoy:1970gd}. On repulsive potentials atom pairs can be excited to a continuum of states. 
We are interested in the dispersive effects of all these shifted optical resonances. 
Writing the additional Faraday rotation analogous to the simplified Faraday model we obtain\footnote{F. Kaminski, N. Kampel, A. Griesmaier, E. Polzik, and J{\"o}rg H. M{\"u}ller, to be published}

\begin{equation}
\theta_F^{pa}(\Delta_A) = \frac{1}{2} \sum_{i,\mathrm{v}} p^{(i)} \frac{\Delta_{\mathrm{v}}^{\mathrm{(i)}}}{\Gamma_{\mathrm{v}}} \int_{-\infty}^{\infty} \sigma^{(i,\mathrm{v})}_{pa}(\vec{r},\Delta_A) \rho(\vec{r}) \mathrm{d}z .
\label{eq:FarMol}
\end{equation}

Here $\Delta_\mathrm{v}^{\mathrm{(i)}}$ is the detuning to the molecular resonance $\mathrm{v}$ for a specific excited state $i$,  $\Gamma_{\mathrm{v}}$ is the linewidth of that resonance. The pair absorption cross section $\sigma^{(i,\mathrm{v})}_{pa}(\vec{r},\Delta_A)$ depends on the density of pairs at a given detuning $\Delta_A$. 
Via the definition $\sigma_{scat} = \Phi_{scat} \frac{\hbar \omega}{I^{inc}} $, where $I^{inc}$ is the incident intensity and $\hbar\omega$ is the energy of a photon, we can relate the pair absorption cross section to the scattered photon flux $\Phi_{scat}$. The scattered photon flux is identical to the atom pair loss rates given by Julienne et al. \cite{Julienne:1996wc,Burnett:1996bq}, since exactly one photon is scattered per atom pair loss event (see App. \ref{A:Mol}). 

The pair absorption cross section $\sigma^{(i,v)}_{pa}$ is linear in density (App. \ref{A:Mol}) and therefore the light assisted cold collision Faraday angle is quadratic in density. Thus the spatial profile of the pair contribution to the Faraday angle has a reduced Gaussian width $w_r^{pa}=w_r/\sqrt{2}$.

We calculate the total Faraday rotation angle by using equations \ref{eq:FarAt} and \ref{eq:FarMol}. To correct for magnetic field inhomogeneities and tensor evolution we multiply the result by the above defined $\beta_B$, such that

\begin{equation}
\theta_F = \beta_B (\theta_F^A + \theta_F^{pa}).
\end{equation}

\begin{figure}
\includegraphics[width=0.5\textwidth]{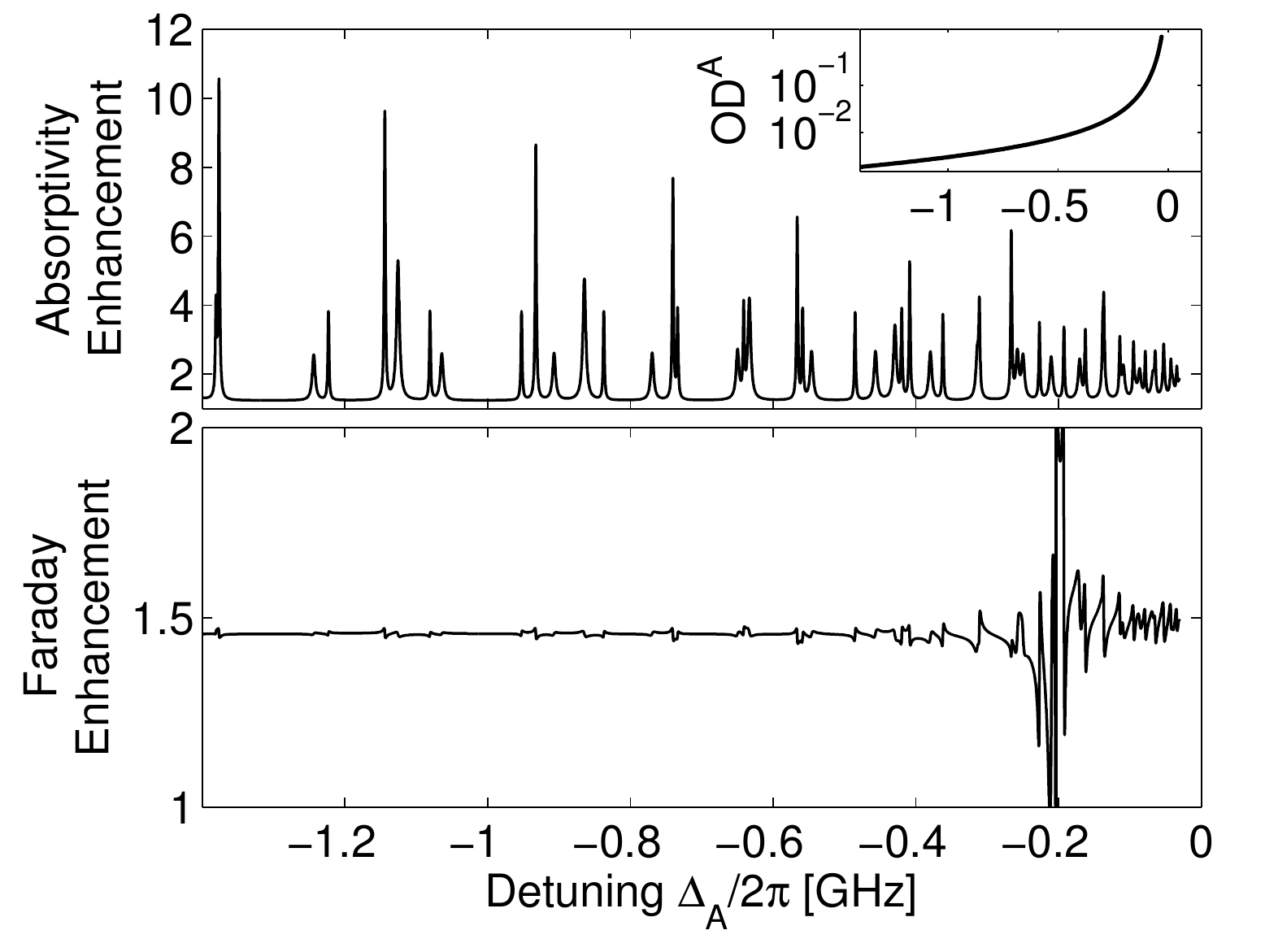}%
\caption{Calculated absorptivity and Faraday angle enhancement due to interactions between atoms. Plotted is $\left( OD^{A}+OD^{pa} \right)/OD^{A}$ and $\left( \theta_F^{A}+\theta_F^{pa} \right)/\theta_F^{A}$ for an atomic density of $f_{opt} \rho_{0}^{abs}=2.6\cdot10^{19}\mathrm{m}^{-3}$. The inset shows the detuning and hyperfine structure corrected independent atom optical depth.}
\label{fig:opticaldepth}
\end{figure}

The total optical depth $OD( \Delta_A )$, defined via the intensity attenuation $ I/I_0 = \exp \left( -OD(\Delta_A) \right) $ is given by the product of the atomic density and the scattering cross section integrated along the propagation direction of the light:

\begin{align}
OD(\vec{r}_{\perp},\Delta_A) &= \int \rho(\vec{r}) \sum_{i,v} \left( \sigma_A^{(i)}(\vec{r},\Delta_A) + \sigma^{(i,\mathrm{v})}_{pa}(\vec{r},\Delta_A) \right) \mathrm{d}z \nonumber \\
&=  OD^{A}(\vec{r}_{\perp},\Delta_A) +  OD^{\mathrm{pa}}(\vec{r}_{\perp},\Delta_A).
\end{align} 

From the determined atom number $f_{opt} N_{\mathrm{at}}^{\mathrm{abs}}$ and ensemble size $w_z$ we calculate the independent atom on-resonant OD for the $|1,-1\rangle$ to $|2,-2\rangle$ D1 transition with $\xi_i^2=1/2$. We find a peak optical depth $OD^{A} = \xi_i^2 (3\lambda_{D1}^2/2\pi) \rho_0 \sqrt{2 \pi} w_z = 680$. 

We present the predicted enhancement due to light assisted cold collisions of absorptivity $(OD^{A}+OD^{pa})/OD^A$ and Faraday angle $(\theta^A+\theta^{pa})/\theta^A$ for the determined atom number as a function of probe detuning $\Delta_A$  in Fig.~\ref{fig:opticaldepth}. The inset gives the independent atom OD, taking into account all relevant levels for Faraday rotation (see Fig.~\ref{fig:setup}).
The enhancement $(OD^{A}+OD^{pa})/OD^{A}$ of the absorptivity due to atomic interactions is about 25\% in between resonances and even up to a factor of 10 close to vibrational resonances in the plotted detuning range. For the Faraday angle we find an almost constant enhancement of 1.46 at all detunings away from $\Delta_0$. The local enhancement due to single vibrational resonances is small.

\section{Discussion and Conclusion}

The data presented in Fig.~\ref{fig:ftrace}, \ref{fig:transmission} and \ref{fig:images} show that the influence of diffraction on Faraday images is reduced. 
Common-mode diffraction on both images, $I_V$ and $I_H$, is canceled when calculating the difference $I_H-I_V$. The common-mode diffraction stems from the scalar polarizability $\alpha^{(0)}$, which is the largest contribution to the polarizability as shown in Fig.~\ref{fig:polarizability}. 
This compensation is, however, not perfect and we discuss in the following the effect of uncompensated diffraction and refraction on the detected Faraday angle.
In a geometric optics picture, the trajectories of light rays are curved due to the inhomogeneous density of the sample. 
This leads for strong refraction to a breakdown of the column density approximation, which implicitly assumes a straight line ray path. Eventually this leads to noticeable differences between the column density and the spatial profile of the Faraday angle.
For red detunings the extended atomic cloud acts as a thick collimating lens, such that ray trajectories are bent towards the center of the cloud, which leads to a reduced Faraday angle in the center.

Differential diffraction and lensing is associated with the $\alpha^{(1)}$ part of the polarizability. It leads to a mismatch in wavefront of the left- and right-handed circular polarization modes at the exit plane of the atomic ensemble. This introduces locally ellipticity to the initially pure linear polarization. 
Since detection in the H/V basis is insensitive to circular polarization this lowers the Faraday angle by a second order correction.

While residual diffraction and refraction reduces the Faraday angle, Fig.~\ref{fig:fdetuning} shows measured peak Faraday angles significantly above the prediction from the Maxwell-Bloch simulation, which assumes the atoms to be independent scatterers and does not include diffraction.
We match our light assisted collision model to the data by choosing $f_{opt}=2.13 < f_{max}$, scaling our inferred atom number from absorption imaging to $f_{opt} N_{\mathrm{at}}^{\mathrm{abs}} = 1.73 \, 10^6$. Comparing the cold collision model for this input atom number to the corresponding Maxwell-Bloch simulation we find an increase of the Faraday angle of 1.46 as indicated in Fig.~\ref{fig:ftrace}. 
Trying to fit the data directly with the Maxwell-Bloch model would require an atom number scaling of $1.46 f_{opt}=3.1$. 
This lies 15\% above the conservative upper bound of $f_{max}=2.7$, discussed at the end of Sec.~\ref{sec:setup} and strongly suggests that the independent scatterer assumption breaks down.

Our strategy to correct the optical response of the gas by considering atom-atom interaction on molecular potentials and the corresponding redistribution of oscillator strength in frequency space can be contrasted to other approaches to describe the optical properties of a dense gas. Instead of calculating the collective response by a systematic expansion in atomic density \cite{Morice:1995gf} or by a configuration average over many randomly placed interacting point dipoles \cite{Sokolov:2009kl}, we focus on the contribution of close pairs for which big resonance shifts occur which lead to important modifications in the wing of the atomic line. The number of close pairs is determined using the quantum mechanical scattering wave function for atom pairs interacting on the ground state molecular potential, hence particle correlations are accounted for. Close to the unperturbed atomic resonance we expect our approach to fail, since for the corresponding large internuclear distances pairs can no longer be considered isolated and the collective response of ever bigger clusters of atoms should be calculated instead. 
Our model predicts a surprisingly large modification of the Faraday rotation angle even for the modest particle density $\rho_0^{abs} f_{opt} \lambdabar^3 = 0.05$ used in the experiment.
The model neglects hyperfine recoupling on molecular potentials and does not explain the observed shift in the position of $\Delta_0$.
Interestingly, in a recent experiment, which employs resonant absorption imaging as a detection method for high density 2-D quantum gases \cite{Rath:2010cz}, a decrease of absorptivity with increasing density has been observed. This is consistent with our simple picture of redistribution of oscillator strength from the line center into the wings due to the resonant dipole-dipole interaction. 

We turn to the suitability of our atomic samples to multimode quantum memories. With the favorable optical depth and Fresnel number a mode capacity in the hundreds is predicted in forward read-out \cite{GrodeckaGrad:2011ug}. In experimental implementations this number will likely be limited by the finite resolution of the imaging system.

While the increased Faraday angle signals a higher coupling between atoms and light it remains to be seen in future experiments to what extend decoherence is increased by the resonant dipole-dipole interaction. For this, the presented weakly destructive dual-port detection method will be an invaluable tool since details of the radial spin density distribution can be examined repeatedly in face of strong refraction and diffraction effects.

\section*{Acknowledgements}
We thank EU projects HIDEAS and EMALI for supporting this project and R. Le Targat for contributing to this work at the early stages of the experiment. We are grateful to A. Grodecka-Grad for supporting simulations of 3D diffraction and to J. Dalibard for sharing related results with us.

\appendix

%
%
%


%

%
%
\section{Light-Matter Interface Simulation \label{sec:lightmatter}}

The interaction Hamiltonian in continuous variables for light propagation along the quantization axis is:

\begin{align}
\mathcal{H}_{int}^{(0)} =& \frac{2}{3} H_0 \alpha^{(0)}(\Delta_A) \hat{S}_0 \mathbb{1} \\
\mathcal{H}_{int}^{(1)} =& H_0 \alpha^{(1)}(\Delta_A) \hat{S}_3 \hat{F}_z \\
\mathcal{H}_{int}^{(2)} =& H_0 \alpha^{(2)}(\Delta_A) \times \nonumber \\
& \Big[ \hat{S}_1 \left( \hat{F}_x^2 - \hat{F}_y^2 \right) + \hat{S}_2 \left( \hat{F}_x \hat{F}_y + \hat{F}_y \hat{F}_x \right) - \nonumber \\
& \hat{S}_0 \left( \hat{F}_z^2 - \tfrac{1}{3} F(F+1) \mathbb{1} \right) \Big],
\end{align}

where we defined 

\begin{equation}
H_0 = \rho A \mathrm{d}z \left( \frac{2 |\langle J|| d_{A} || J' \rangle|^2 }{\hbar \Gamma_A} \right) \left( \frac{\hbar \omega}{2 \epsilon_0 A} \right)
\end{equation}

and omitted the space and time (z,t)-dependence of all operators. A is the interaction area, $d_A$ is the transition dipole moment and $\epsilon_0$ is the vacuum permittivity.

The explicit commutation relations for continuous variables are:

\begin{align}
\left[\hat{S}_i(z,t),\hat{S}_j(z',t') \right] &= i \sum_k \epsilon_{ijk} \hat{S}_k(z,t) \delta(z-z') \delta(t-t') \\
\left[ \hat{f}_i (z,t) ,\hat{f}_j(z',t') \right] &= \frac{i}{\rho A} \sum_k \epsilon_{ijk} \hat{f}_k(z,t) \delta(z-z') \delta(t-t') 
\label{eq:ContComRel}
\end{align}

and the Stokes operators with units $[\hat{S}]=[N_{ph}/cT]=1/\mathrm{meter}$, number of photons $N_{ph}$ over speed of light c and interaction time T are defined as, again omitting (z,t):

\begin{align}
\hat{S}_0 = & \frac{1}{2} \left( \hat{a}_+^\dag \hat{a}_+ +  \hat{a}_-^\dag \hat{a}_- \right) & \label{eq:Stokes0}
\\
\hat{S}_1 = & \frac{1}{2} \left( \hat{a}_H^\dag \hat{a}_H -  \hat{a}_V^\dag \hat{a}_V \right) \label{eq:Stokes1}
\\
\hat{S}_2 = & \frac{1}{2} \left( \hat{a}_{45}^\dag \hat{a}_{45} -  \hat{a}_{-45}^\dag \hat{a}_{-45} \right) \label{eq:Stokes2}
\\
\hat{S}_3 = & \frac{1}{2} \left( \hat{a}_+^\dag \hat{a}_+ -  \hat{a}_-^\dag \hat{a}_- \right). \label{eq:Stokes3}
\end{align}

We solve the equations of motion for the F=1 manifold with all three ground state levels. We choose to propagate the density matrix $\hat{\sigma}$ and the second order coherence matrix for the light $\hat{p}=(\hat{S}_1+\hat{S}_2+\hat{S}_3)/\hbar$:

\begin{align}
\frac{\partial \hat{p}}{\partial z} =& i \left[\hat{p},\mathcal{H}_{int}+\mathcal{H}_B \right] \\
\frac{\partial \hat{\sigma}}{\partial t} =& i \left[ \hat{\sigma},\mathcal{H}_{int}+\mathcal{H}_B \right] .
\end{align}

The Stokes operators are related to the experimentally taken images by $S_1 \propto I_H-I_V$ and $S_0 \propto I_H+I_V$. The Faraday angle $\theta_F$ is the rotation angle of linearly polarized light and is thus half the angle of the Stokes vector rotation in the $S_1$/$S_2$ plane of the Poincar\'{e} sphere and reads in Stokes parameters

\begin{equation}
\theta_F=
\frac{1}{2}\arcsin \left( \frac{S_1}{S_0} \right).
\end{equation}

\section{Light Assisted Cold Collision Parameters \label{A:Mol}}

The partial atom loss rate due to photoassociation to the level $(i,\mathrm{v})$ is given by \cite{Julienne:1996wc,Burnett:1996bq}

\begin{equation}
\gamma^{(i,\mathrm{v})}_{pa}(\vec{r},\Delta_A) = 2 \rho(\vec{r}) K_{\mathrm{v}} \xi_i^2 \frac{ \nu_{\mathrm{v}} \Gamma_{\mathrm{v}} }{ \Delta^{(i)2}_{\mathrm{v}} + (\Gamma_{\mathrm{v}}/2)^2 },
\end{equation}

where $\Gamma_{\mathrm{v}}$ is the full linewidth of the molecular resonance $\mathrm{v}$ and $\Delta^{(i)}_{\mathrm{v}}$ is the detuning relative to this resonance. The parameter $\nu_{\mathrm{v}}$ characterizes the slope of the excited state molecular potential 
in frequency units and $K_{\mathrm{v}}$ is the rate coefficient.


As explained in Sec. \ref{sec:models} the partial pair absorption cross section is then given by

\begin{equation}
\sigma^{(i,\mathrm{v})}_{pa}(\vec{r},\Delta_A) = \frac{\gamma_{pa}^{(i,\mathrm{v})}(\vec{r},\Delta_A)}{2} \frac{\hbar \omega}{I^{inc}}.
\end{equation}

\bibliography{Papers}
\bibliographystyle{unsrt}

\end{document}